# Spherical Plasmonic Heterodimers: Reversal of Optical Binding Force as the Effect of Symmetry Breaking

M.R.C. Mahdy*[1,2,3], Md. Danesh[2], Tianhang Zhang[2,4], Weiqiang Ding[5], Ariful Bari Chowdhury[6], Hamim Mahmud Rivy[1], MQ Mehmood[2,7]

[1]*Department of Electrical & Computer Engineering, North South University, Bashundhara, Dhaka 1229, Bangladesh*

[2]*Department of Electrical and Computer Engineering, National University of Singapore, 4 Engineering Drive 3, Singapore 117583*

[3]*Pi Labs Bangladesh Ltd., ARA Bhaban, 39, Kazi Nazrul Islam Avenue, Kawran Bazar, Dhaka, Bangladesh*

[4]*NUS Graduate School for Integrative Sciences and Engineering, National University of Singapore, 28 Medical Drive, Singapore 117456*

[5]*Department of Physics, Harbin Institute of Technology, Harbin 150001, People's Republic of China*

[6]*Department of Public Health, North South University, Bashundhara, Dhaka 1229, Bangladesh*

[7]*Department of Electrical Engineering, Information Technology University of the Punjab, 54000 Lahore, Pakistan*

[*] Corresponding author: mahdy.chowdhury@northsouth.edu

**Abstract**

**The stimulating connection between the reversal of near field plasmonic binding force and the role of symmetry breaking has not been investigated in detail in literature. As both bonding and anti-bonding modes are present in the visible spectra of well-known spherical plasmonic heterodimer sets, binding force reversal is commonly believed to occur for all such heterodimers. But our work suggests a very different proposal. We demonstrate that for the symmetry broken heterodimer configurations: reversal of lateral (for on-axis heterodimers) and longitudinal (for off-axis heterodimers: end-fire and nearly end-fire configurations) near field binding force follow fully distinct mechanisms; i.e. later one on relative orientation and constructive dipole-quadrupole resonance but the former one on light polarization and the induced electric resonance. Interestingly, the reversal of longitudinal near field binding force can be easily controlled just by changing the direction of light propagation or just their relative orientation. Though it is commonly believed that plasmonic forces mostly arise from the surface force and Fano resonance can be a promising way to achieve binding force reversal; our study based on Lorentz force dynamics suggests notably opposite proposals (for both instances) for the case of plasmonic spherical heterodimers. Observations of this article can be very useful for improved sensors, particle clustering and aggregation.**

Fano resonances, super-scattering and plasmonic hybridization in nanostructures [1-4] have received substantial attention in the area of plasmonics. The promising applications of plasmonic hybridization, super-scattering and Fano resonances have been investigated in improved sensitivity of the resonance [5], bio sensing [6], surface-enhanced Raman scattering [7,8], photonic propagation and wave guiding [9,10], plasmon-induced transparency [11] to super scattering [12] and many others [13,14]. In contrast, much less attention is dedicated on near field optical force due to Fano resonance and plasmonic hybridization; especially for plasmonic dimers [1,2,15-19]. When two metal nanoparticles are placed very close to each other, the properties of their surface plasmons are dramatically modified. This configuration of nanoparticles is known as a "dimer". Behaviors of such dimers have been studied in refs. [1,2,4,15-19]. Specially heterodimers show remarkable properties such as Fano resonances [1,17], avoided crossing behavior [1], optical nanodiode effect [1] and so on. But the behavior of near field optical force for such heterodimers have not been studied in detail. So far only two works [20, 21], as far as of our knowledge, have studied the behavior of binding force for *on-axis spherical heterodimers*.

Though the behavior and reversal of near field optical binding force for spherical plasmonic on-axis *homodimers* [22- 26] (due to bonding and anti-bonding modes without any substrate) have been studied comprehensively but such detailed investigations lack for on-axis [20, 21] and *off-axis spherical heterodimers*. Here off-axis means end-fire [20] and nearly end-fire configuration [cf. Fig. 1 when the rotation angle, $\varphi$, of the particle is between 60 to 120 degrees]. Considering the heterodimer cases, based on Fano resonance, the reversal of near field optical binding force has been reported in [27] and [28] for nanorod structures [27] and for disk along with a *ring* structure [28]. The answer of the question, whether such Fano resonance (raised from heterodimer interaction) is a universal process of achieving binding force reversal or not, is still unknown.

When the incident light field is polarized along the dimer axis (longitudinal polarization), the hybridized in-phase combination reflects a bonding mode $\sigma$ and the out-of-phase mode represents an antibonding $\sigma^*$ configuration, respectively. However, when the polarization is perpendicular to the inter particle axis (transverse polarization), the scenario is exactly reversed: the in-phase combination is an antibonding mode ($\pi^*$) and the out-of-phase combination is a bonding mode ($\pi$) [2]. A homodimer structure under longitudinal polarization supports only bonding plasmon mode ($\sigma$). Likewise, only a $\pi^*$-mode is observed in a homodimer under transverse polarization. But a plasmonic "heterodimer" structure supports both bonding and antibonding plasmon modes at the same time due to its broken symmetry [2]. Hence it is expected that binding force reversal should occur almost for all the spherical heterodimer structures.

But our work suggests that reversal of lateral (on-axis heterodimers) and longitudinal (off-axis heterodimers) binding force of symmetry broken heterodimers follow fully different key parameters/mechanisms. Most importantly, the reversal of longitudinal binding force can be easily controlled due to forced symmetry breaking just by changing the direction of wave propagation for a specific set-up of off-axis heterodimers or by changing their relative orientation. Interestingly, though reversal of optical binding force occurs for nano rods or other shapes due to Fano resonance [27,28], we have demonstrated that Fano resonance does not contribute to binding force reversal for spherical heterodimers [1,4]. These observations are quite different than the homo-dimer cases reported in [22-26].

Though Lorentz force analysis has been applied previously in [29-32] to understand the mechanism of chirality induced force, Luneburg lenses, mechanical interaction between light and graded index media, cloaking effect, background effect on radiation pressure; such analysis is rare to understand the plasmonic effects and plasmonic binding force. Although it is commonly believed that plasmonic forces mostly arise from the surface force/polarization induced charges [33, 34], our study suggests a notably different proposal especially for the off-axis heterodimers.

*Table-1 of this article (given below) represents a very short overview of our overall investigation throughout this article.* However, Table-1 represents only few possible cases which ultimately lead us to the possible final two conclusions of this article (those two final conclusions are demonstrated in detail in Fig. 2):

'Reversal of optical longitudinal binding force can be easily controlled by manually controlling the relative orientation (in two distinct ways) of the spherical heterodimers as shown in detail in Fig. 2 (and it will be discussed again in forthcoming last section just before the CONCLUSION section).'

Observations of this article for spherical dimers can be very useful for the future plasmonic applications of the heterodimers (even with the cubic or other shapes, when the symmetry is broken) in the areas of improved sensors [4-6], particle clustering and aggregation [22-24].

**Table 1: An overview on the behavior of binding force for on and off-axis heterodimers (only few possible cases)**

| Heterodimer Set | Number in Fig.1 | On-axis [$\varphi=0$] | Off-Axis [$\varphi=60$ to $120$ deg.] | Pol. | Binding Force reversal: (i)Lateral (ii)Longitudinal | Comment (a) Inter-particle edge to edge gap, d, is always fixed 20 nm. (b) Heterodimer radii are fixed: 50 nm and 100 nm. |
|---|---|---|---|---|---|---|
| Ag-Au | (a) [=(d)] | Yes | | $\perp$ | (i)Yes | Lateral binding force reverses only for perpendicular polarization. For higher wavelength region: Reversal of force occurs due to zero surface and bulk force at a specific wavelength near bonding (attractive force) resonance. For lower wavelength region: such force reversal can be recognized from the reversal of electric dipole moment of the smaller object. In fact, such reversals (repulsive to attractive) occur due to induced electric resonance near the bonding resonance mode. |
| Au-Au | (b) [=(e)] | Yes | | $\perp$ | (i)Yes | |
| Ag-Ag | (c) [=(f)] | Yes | | $\perp$ | (i)Yes | |
| Ag-Au | (a) [=(d)] | Yes | | $\parallel$ | (i)No | |
| Au-Au | (b) [=(e)] | Yes | | $\parallel$ | (i)No | |
| Ag-Ag | (c) [=(f)] | Yes | | $\parallel$ | (i)No | |
| Ag-Au | (a) | | Yes | $\perp$ and $\parallel$ | (ii) No | Longitudinal binding force reverses [for only Ag-Au and Ag-Ag case] only when the bigger particle rotates and the light is perturbed by the fixed smaller object at first. This reversal occurs due to the constructive dipole-quadrupole resonance and due to the dominance of the bulk Lorentz force. However, for all heterodimer sets attractive and repulsive force can be very easily controlled by changing the light propagation direction or changing the relative orientation of the dimers. Such control is not possible with the spherical homo-dimers. |
| Au-Au | (b) | | Yes | $\perp$ and $\parallel$ | (ii) No | |
| Ag-Ag | (c) | | Yes | $\perp$ and $\parallel$ | (ii) No | |
| Ag-Au | (d) | | Yes | $\perp$ and $\parallel$ | (ii) Yes | |
| Au-Au | (e) | | Yes | $\perp$ and $\parallel$ | (ii) No | |
| Ag-Ag | (f) | | Yes | $\perp$ and $\parallel$ | (ii) Yes | |

**RESULTS AND DISCUSSION**

We specify that throughout this paper we refer to 'exterior' or 'outside' forces as those evaluated outside the volume of the macroscopic particles, while 'interior' or 'inside' refer to those quantities inside this object volume. To consider the realistic effects, we have done all the numerical calculations /full wave simulations [35] in three dimensional (3D) structures.

The proposed simplest set-up is illustrated in Fig. 1. The Gold and Silver particles are placed near to each other. The real and imaginary part of the permittivity of Gold and Silver are taken from the standard CRC and Palik data [35-37]. Inter particle distance is 'd'. The source is a simple *x*-polarized plane wave $E_x = E_0 e^{i\beta y}$ propagating in '-*y*' direction. This set-up is a 'forced' symmetry broken system, which later plays a vital role for the force reversal. If the heterodimer set-up is shined from the top, such 'forced' symmetry breaking is not possible. The 'outside optical force' [38, 39] is calculated by the integration of time averaged Minkowski [20-28, 38, 39] stress tensor at r=$a^+$ employing the background fields of the scatterer of radius *a*:

$$\langle \boldsymbol{F}_{\text{Total}}^{\text{Out}} \rangle = \oint \langle \bar{\bar{\boldsymbol{T}}}^{\text{out}} \rangle \cdot d\boldsymbol{s}$$
$$\langle \bar{\bar{\boldsymbol{T}}}^{\text{out}} \rangle = \frac{1}{2}\text{Re}[\boldsymbol{D}_{\text{out}}\boldsymbol{E}_{\text{out}}^* + \boldsymbol{B}_{\text{out}}\boldsymbol{H}_{\text{out}}^* - \frac{1}{2}\bar{\bar{\boldsymbol{I}}}(\boldsymbol{E}_{\text{out}}^* \cdot \boldsymbol{D}_{\text{out}} + \boldsymbol{H}_{\text{out}}^* \cdot \boldsymbol{B}_{\text{out}})]$$  (1)

Where 'out' represents the exterior total field of the scatterer; *E*, *D*, *H* and *B* are the electric field, displacement vector, magnetic field and induction vectors respectively, $\langle \rangle$ represents the time average and $\bar{\bar{\boldsymbol{I}}}$ is the unity tensor.

On the other hand, based on the Lorentz force, the total force (surface force and the bulk force [29-32]) can be written as:

$$\langle \boldsymbol{F}_{\text{Total}} \rangle = \langle \boldsymbol{F}_{\text{Volume}} \rangle = \langle \boldsymbol{F}_{\text{Bulk}} \rangle + \langle \boldsymbol{F}_{\text{Surf}} \rangle = \int \langle \boldsymbol{f}_{\text{Bulk}} \rangle dv + \int \langle \boldsymbol{f}_{\text{Surface}} \rangle ds$$  (2)

Where

$$\langle \boldsymbol{f}_{\text{Surface}} \rangle = \sigma_e \boldsymbol{E}^*_{avg} + \sigma_m \boldsymbol{H}^*_{avg}$$
$$= \{\epsilon_o (\boldsymbol{E}_{\text{out}} - \boldsymbol{E}_{\text{in}}) \cdot \hat{\boldsymbol{n}}\} \left(\frac{\boldsymbol{E}_{\text{out}} + \boldsymbol{E}_{\text{in}}}{2}\right)^* + \{\mu_0 (\boldsymbol{H}_{\text{out}} - \boldsymbol{H}_{\text{in}}) \cdot \hat{\boldsymbol{n}}\} \left(\frac{\boldsymbol{H}_{\text{out}} + \boldsymbol{H}_{\text{in}}}{2}\right)^*, \quad (3)$$

$$\langle \boldsymbol{f}_{\text{Bulk}} \rangle = \frac{1}{2}\text{Re}[\varepsilon_0 (\nabla \cdot \boldsymbol{E}_{\text{in}}) \boldsymbol{E}^*_{\text{in}} + \mu_0 (\nabla \cdot \boldsymbol{H}) \boldsymbol{H}^*_{\text{in}}] - \frac{1}{2}\text{Re}[i\omega(\varepsilon_s - \varepsilon_0)\{\boldsymbol{E}_{\text{in}} \times \boldsymbol{B}^*_{\text{in}}\} + i\omega(\mu_s - \mu_0)\{\boldsymbol{D}^*_{\text{in}} \times \boldsymbol{H}_{\text{in}}\}] \quad (4)$$

$\boldsymbol{f}_{\text{Surface}}$ is the surface force density (the force which is felt by the bound electric and magnetic surface charges of a scatterer), which is calculated just at the boundary of a scatterer [29-32]. $\boldsymbol{f}_{\text{Bulk}}$ is the bulk force density, which is calculated from the interior of the scatterer by employing the inside field [29-32]. 'in' represents the interior fields of the scatterer; 'avg' represents the average of the field. $\sigma_e$ and $\sigma_m$ are the bound electric and magnetic surface charge densities of the scatterer respectively. $\varepsilon_s$ is permittivity and $\mu_s$ is permeability of the scatterer. The unit vector $\hat{\boldsymbol{n}}$ is an outward pointing normal to the surface. As long as we know, the Lorentz force dynamics of plasmonic particles and especially heterodimers have not been discussed previously. It is notable that the 'external dipolar force' [38, 39] (which has also been described as Lorentz force in [20]) is quite different than the Lorentz force defined in our Eqs (2) - (4). Even if the quasi static analysis (i.e. dipolar force [40,41]) leads to wrong conclusion (for example- in refs [20, 42-44]); the agreement of Lorentz volume force [29-32,45] and external ST method [38, 39, 46-50] based on full electrodynamic analysis, which is considered for all the force calculations in this article, should lead to the consistent result for realistic experiments.

**Lateral binding force: On-Axis Spherical Heterodimers**

Behavior of optical binding force for on-axis spherical heterodimers has been studied in [20] considering the inter particle edge to edge gap of only 2nm. In addition, the size of the spherical objects has been considered only 10 nm and another one maximum 40 nm in [20]. However, we have observed that if the inter particle gap is increased (i.e. 20 nm instead of 2 nm), the reversal of optical binding force dies out for both polarizations of light. Still by optimizing the size of the heterodimers a more generic way of binding force reversal has been demonstrated in the next sub-sections.

(a) **Parallel Polarization: No reversal of lateral binding force for Au-Ag, Au-Au and Ag-Ag on-axis heterodimers**

For Ag-Au, Au-Au and Ag-Ag heterodimer configuration, the lateral binding force, $F_{Bind\,(x)} = (F_{B\,(x)} - F_{S\,(x)})$, reversal does not occur for the light polarized parallel to the dimer axis [cf. Fig. 2s in Supplement S2 (a) and Fig. 5s (c) in supplement S3]. Here $F_{B(x)}$ and $F_{S(x)}$ are the +$x$-directed time averaged force on big and small particle respectively. According to our discussion in Supplement S2(a), the important conclusion is that although reversal of lateral optical binding force occurs for nano rods or other shapes due to Fano resonance [27, 28], Fano resonance is in general not the reason of the reversal of optical binding force.

(b) **Perpendicular Polarization: Reversal of lateral binding force for Au-Ag, Au-Au and Ag-Ag on-axis heterodimers**

At first, we consider two on-axis Ag-Au, Au-Au and Ag-Ag particles of 100 and 50 nm with inter particle distance (edge to edge distance) of 20 nm [cf. Figs. 1 (a), (b) and (c)] and perpendicular polarized light.

(1) It is observed that near the bonding resonance mode reversal of the optical binding force (negative to positive) occurs at the wavelength of 646 nm [cf. Fig. 3(a), (b) for Ag-Au and Fig. 3 (e), (f) for Au-Au]. It should be noted that: $F_{Bind\,(x)} = \left(F_{B\,(x)} - F_{S\,(x)}\right) = \text{Del } F_{Bulk(x)} + \text{Del } F_{Surf(x)}$; which is defined in Supplement S2 (b). Reversal of optical binding force occurs at that specific wavelength mainly due to the individual zero surface ($\text{Del } F_{Surf\,(x)} = 0$) and bulk ($\text{Del } F_{Bulk(x)} = 0$) Lorentz force [cf. Fig. 3 (c), (d) for Ag-Au and Fig. 3(g), (h) for Au-Au; detail analysis is given in Supplement S2(b)]. A sudden change is observed in the phase of steady current as well as in the surface charge distribution in Fig. 3s in Supplement S2 (b).

(2) It is also demonstrated that whenever the 2$^{nd}$ reversal (positive to negative) of the lateral binding force occurs for transverse/perpendicular polarization near the wavelength 500 nm [cf. Fig. 3(a), (b) for Ag-Au and Fig. 3 (e), (f)], the real part of the induced electric dipole moment reverses its sign near the resonance of the smaller object in Supplement S2 (b) [which does not occur for parallel polarized case]. So, the reversal of lateral binding force near this specific wavelength can better be explained based on the idea of induced same or opposite electric charges similar to the idea (reversal of the electric polarizability near resonance) proposed in ref. [20]. Results of on-axis Ag-Ag heterodimers are like Ag-Au and Au-Au cases; which have been demonstrated in supplement S3.

**Longitudinal binding force for Off-Axis Plasmonic Heterodimers: end-fire and near end-fire configuration**

In this section, we mainly focus on the Ag-Au heterodimers to explain the behaviour of the off-axis heterodimers. Reversal of the optical binding force has been observed for only Ag-Au and Au-Au off-axis heterodimers (only when the smaller object perturbs the propagating light at first) and this issue relates to the presence of constructive interference of dipole-quadrupole mode. Another notable point is that: mutual attraction and repulsion of all the off-axis heterodimers can be easily controlled by changing the direction of propagating light or by changing the relative orientation of the particles. All the conclusions of the forthcoming sections have been noted very shortly in Table-1.

**Au-Ag off-axis heterodimers: Longitudinal binding force for both polarizations**

Now, we consider Au-Ag particles of 100 and 50 nm respectively with inter particle distance of 20 nm [cf. Fig.1 (a) and (d)] but considering that the rotation angle, $\varphi$, of the particle is between 60 to 120 degrees [i.e. end fire or nearly end fire configuration [20]]. The source is same. We start to create angular displacement from the $x$- axis considering two cases: (A) Rotating the smaller object and fixing the bigger one [cf. Fig. 1(a)] and (B) Rotating the bigger object while keeping the smaller one fixed [cf. Fig. 1(d)]. Now the question arises: 'Should there be any difference on longitudinal optical binding force for these two cases- (A) and (B)?'. The notable observation of this work: the behavior of longitudinal binding forces is quite different for these two cases due to the forced breaking of symmetry by placing the light source at one side of the dimer configuration instead of at the top of the set-up. If the light source were placed at the top of the set-up, such difference should not arise. According to our forthcoming observations, forced symmetry breaking is detected as one of the key ways to control the inter-particle attraction and repulsion. Some previous symmetry broken set-ups have been discussed in [51, 52] (but not for optical force), which are different than our case.

However, for both cases- (A) and (B), the extinction cross sections reveal that bonding mode resonance blue shifts for increasing angular displacement in Fig. 4(a), (c), (e), (g) for both aforementioned cases [also cf. supplement S4 for the case of Au-Au heterodimers with parallel polarization of light; rotating the smaller object]. It appears that an 'angular ruler' may also be possible like previously defined 'inter-particle gap ruler' in ref. [53].

For the off-axis heterodimers, the attractive force can be defined as the positive value of the optical binding force $F_{Bind\,(y)}(SR)= (F_{S\,(y)}-F_{B\,(y)})$ and $F_{Bind\,(y)}(BR)= (F_{B\,(y)}-F_{S\,(y)})$ [here SR means small rotating and BR means big rotating], considering two important facts: (a) the angular displacement angles should be much higher and $\varphi$ should be as close as 90 degree [i.e. $60 < \varphi < 120$] and (b) $x$-directed lateral force $F_{(x)}$ is at least ten times smaller than $y$-directed force $F_{(y)}$ (which is usually satisfied, as the $y$-directed scattering force is usually much higher than the $x$-directed lateral force for plasmonic spherical heterodimers). It should also be noted that the scattering force of the bigger object is always pushing force [negative value of $F_{B\,(y)}$], which is one of the key issues to control the reversal of the $y$-directed binding force (this will be explained next).

(i) **Au-Ag off-axis heterodimers: Rotating the smaller particle and fixing the big one**

At first, we consider the rotation of the smaller object [case (A); cf. Fig. 1(a)] for both perpendicular and parallel polarizations of the light. For $\varphi=60$ to 90 degrees, it is observed that only the scattering force of the smaller object experiences the reversal at bonding resonance region. On the other hand, scattering force of the bigger object ($F_{B\,(y)}$) is always pushing force. But the most important fact is that $|F_{S\,(y)}| < |F_{B\,(y)}|$ ; always. As a result, $F_{Bind\,(y)}(SR) = (F_{S\,(y)}-F_{B\,(y)})$ is always positive [attractive force as shown in Fig. 4(b) and (f)]. Importantly, the real part of electric dipole moment of the smaller object reverses its sign near the bonding resonance [not shown] but $F_{Bind\,(y)}(SR)$ always remains attractive with no reversal of sign. In fact, the difference of the particle radius of both the particles plays a vital role. When one of the particles in the heterodimer is much larger than the other one and the propagating light is perturbed by the bigger object at first, the scattered field from the large particle becomes much larger compared to the incident field. When the field enhancement is quite high at the inter particle gap position, this enhanced field forces the dipole on the small particle to oscillate in phase. Accordingly, with larger radius of bigger particle, the optical force between the particles becomes always attractive.

(ii) **Au-Ag off-axis heterodimers: Rotating the bigger particle while keeping the small one fixed**

Now, we shall consider the alternate orientation [case (B); cf. Fig. 1 (d)] by rotating the bigger object and fixing the smaller object. If the bigger object is rotated and the smaller one is fixed, $F_{Bind(y)}$ (BR) reverses during the antibonding type resonance mode and near spectral dip position. This is explained next.

When the smaller object is rotated and the propagating light is perturbed by the bigger object at first, the scattering force on the bigger object (always pushing) is always higher than the smaller one. But when the bigger object is rotating and the propagating light is perturbed by the smaller object at first, there are some chances to find some wavelength regions when the scattering force on the smaller object becomes higher than the bigger object. In this way, the binding force $F_{Bind(y)}$ (BR)= ($F_{B(y)}$-$F_{S(y)}$) can be observed attractive in those wavelength regions. This is what exactly happens during the anti-bonding type resonance modes as shown in Fig. 4 (c), (d) and (g), (h); which is quite different than the conventional idea of optical binding force with homodimers [24]. For homodimers, according to the quasi-static approximation limit [24]: the bonding modes and antibonding modes have positive and negative definite slopes, respectively. As a result, consequently they must, respectively, induce attraction and repulsion. But we clearly observe the opposite scenario for the heterodimer set (at a fixed edge to edge distance of 20 nm) when the light is perturbed by the smaller object at first. Then the question arises why this kind of opposite behavior is observed for such symmetry broken heterodimer sets. Its answer lies in the electrodynamics calculations and force distribution analysis instead of the quasi-static analysis; mainly due to the generation of multipoles. Based on the results demonstrated in Fig. 5 (a) - (h) we shall discuss the detail dynamics considering a specific case: $\varphi = 60$ degree.

In Fig. 5 (d) and (h) we have plotted the difference of the bulk Lorentz force, which clearly suggests that the total binding force is dominated by the bulk part of Lorentz force [which is in contrast with the commonly observed dominance of surface [33]/polarization charge induced force [34] for plasmonic objects]. This force can be considered as the scattering force part [33, 54] of the total force, which is physically originating from the multiple scattering between the smaller and the bigger object. Fig. 5 (c) and (g) suggest that during the anti-bonding resonance mode, the directive forward scattering of the bigger object is much higher than the smaller object. Surface charge plots in Fig. 6 suggest that for the parallel polarized illumination, during the wavelength spectrum around 350 nm to 470 nm, constructive interference occurs due to dipole quadrupole resonance. Though this is not super-scattering [3], it is recognized that the forward scattering of the bigger object increases during these spectra [cf. the extinction spectra in Fig. 4(g) where the magnitude of extinction co-efficient increases for higher rotation angles during this specific spectrum regime]. On the other hand, exactly opposite scenario takes place for the bonding mode resonance. For example at higher wavelength regime during bonding mode resonance the smaller object even experiences optical pulling force [cf. Fig. 5(c) and (g)] because of: (i) very strong effective forward scattering along with (ii) more reflected light from the bigger object.

It is observed that the reversal wavelength of the optical binding type force $F_{Bind\,(y)}(BR) = (F_{B\,(y)} - F_{S\,(y)})$ remains almost fixed along with the spectral dip position [cf. Fig. 4(c), (d) and (g), (h)], though the bonding mode resonance blue shifts gradually with the rotation of the bigger object. Moreover, very similar to ref [27], the reversal of the phase of the steady state current takes place near the spectral dip in our heterodimer set-ups as shown in Fig. 6 (m)-(r), though it is constructive dipole quadrupole resonance instead of destructive one reported in [27].

**Simplest procedure to reverse the longitudinal binding force for all the off-axis heterodimers**

So far, we have shown that: *when the bigger object is rotated and the propagating light is perturbed by the smaller object at first*, for only Ag-Au and Ag-Ag off-axis heterodimers reversal of longitudinal binding force occurs [i.e. the dynamics of Ag-Ag and Au-Au heterodimers have been discussed in detail in Supplement S5]. Especially after the anti-bonding resonance mode, the longitudinal binding force is observed always *repulsive* for such heterodimers. On the other hand, for Au-Au heterodimers this force is always *repulsive* for such configuration for the visible wavelength spectrum.

In contrast, *when the smaller object is rotated and the propagating light is perturbed by the bigger object at first*, for all the spherical heterodimers no reversal of longitudinal binding force occurs. Binding force is always *attractive* for such configuration.

So, based on the aforementioned simple observations, finally we can conclude: (1) if we consider *the higher wavelength regions* and change the direction of propagating light manually by bringing the light source from one side of the dimers to another side, it will be easily possible to observe the mutual repulsion and attraction of all the heterodimer sets just due to the automatic change of the relative dimer position of smaller and bigger objects [cf. Fig. 2(b)]. Or (2) simply by changing the relative orientation of the heterodimers manually (not light propagation direction), it is also possible to observe such reversal at higher wavelength regions. Such simplest controls of force reversal due to 'forced broken symmetry' are certainly impossible with the plasmonic homo-dimers [cf. Fig. 2(c)].

**CONCLUSIONS**

To identify the conclusions of this article at a glance, we have listed our key observations in Table 1. It is expected that binding force reversal should occur almost for all the spherical heterodimer structures due to the presence of bonding and anti-bonding mode in the visible spectra. But our work suggests that reversal of lateral (for on-axis heterodimers) and longitudinal (for off-axis heterodimers of end-fire or nearly end-fire configurations) binding force of symmetry broken heterodimers follow fully different key parameters/mechanisms; i.e. later one depends on relative orientation and constructive dipole-quadrupole resonance but the former one on light polarization and the induced electric resonance. Most importantly, the reversal of longitudinal binding force can be easily controlled due to forced symmetry breaking just by changing the direction of wave propagation for a specific set-up of off-axis heterodimers or by changing their relative orientation. In addition, though it is commonly believed that plasmonic forces mostly arise from the surface force and Fano resonance can be a promising way to achieve binding force reversal, our study based on Lorentz force dynamics suggests notably opposite proposals (for both) for the case of plasmonic spherical heterodimers. These demonstrations are quite different than the observations of binding force reversal of the homodimers reported in [22-26]. Proposals of this article should be very useful for improved sensors [4-6, 55], particle clustering and aggregation [22-24].

**ACKNOWLEDGEMENTS**

M.R.C. Mahdy acknowledges Associate Professor Qiu Cheng Wei at National University of Singapore for some interesting discussions.

**FIGURE CAPTION LIST**

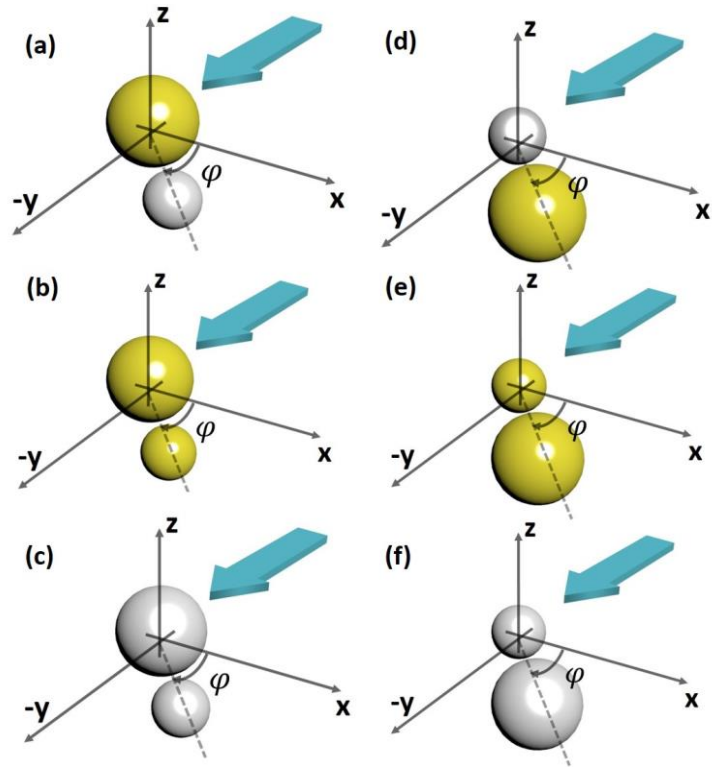

Fig.1: Two particles of radii 100 and 50 nm are placed with inter particle distance from surface to surface '$d$'; $d$= 20 nm throughout the article. One particle centered at $(0,0,0)$ and the other centered at $(R\cos\varphi, -R\sin\varphi, 0)$ with $R = d + r_1 + r_2 = 170$ nm the center-to-center distance of the two object. The angular displacement is '$\varphi$' which is considered 0 degree when the dimers are on-axis in $x$ direction. And the angular displacement is considered +90 degree [end fire configuration] when the dimers are on-axis in $-y$ direction. Two different polarized light sources are applied propagating towards $-y$ direction [in order to break the symmetry; light is shined from a specific side]: (i) For parallel polarization: $x$-polarized plane wave $E_x = E_0 e^{i\beta y}$ (ii) For perpendicular polarization: $z$-polarized plane wave $E_z = E_0 e^{i\beta y}$. Yellow color represents Au and Silver color represents Ag: (a) Au-Ag (b) Au-Au (c) Ag-Ag (d) Ag-Au (e) Au-Au and (f) Ag-Ag.

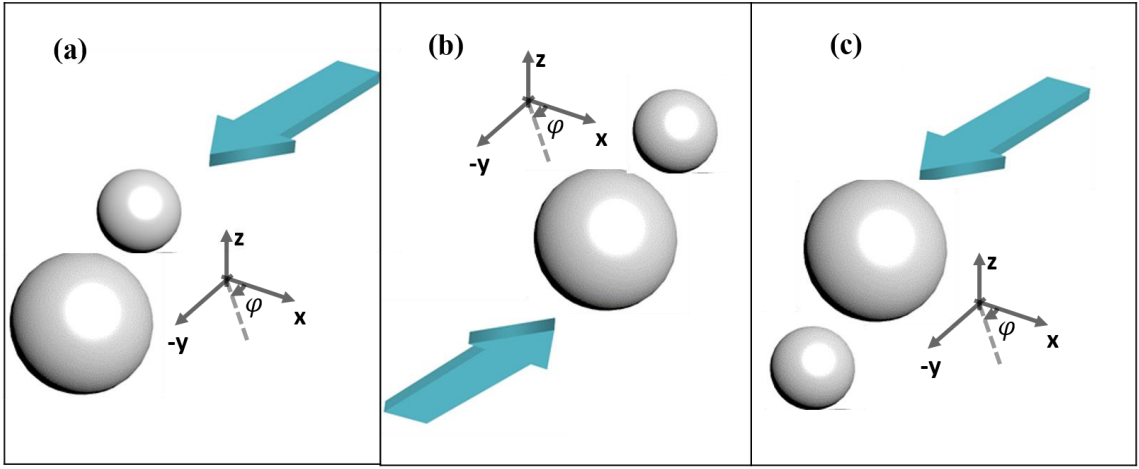

Fig. 2 (a) The configuration of rotating the bigger object by keeping the smaller object fixed as discussed in previous figure (light is propagating towards '-y' direction). (b) For the configuration of Fig (a), the change of the direction of propagating light manually by bringing the light source from one side of the dimers to another side (light is propagating towards '+y' direction). It will be easily possible to observe the mutual repulsion and attraction of all the heterodimer sets just due to the automatic change of the relative dimer position of smaller and bigger objects at higher wavelength regions (discussed in main text before conclusion). (c) However, by changing the relative orientation of the heterodimers manually (not light propagation direction), it is also possible to observe reversal of binding force at higher wavelength regions.

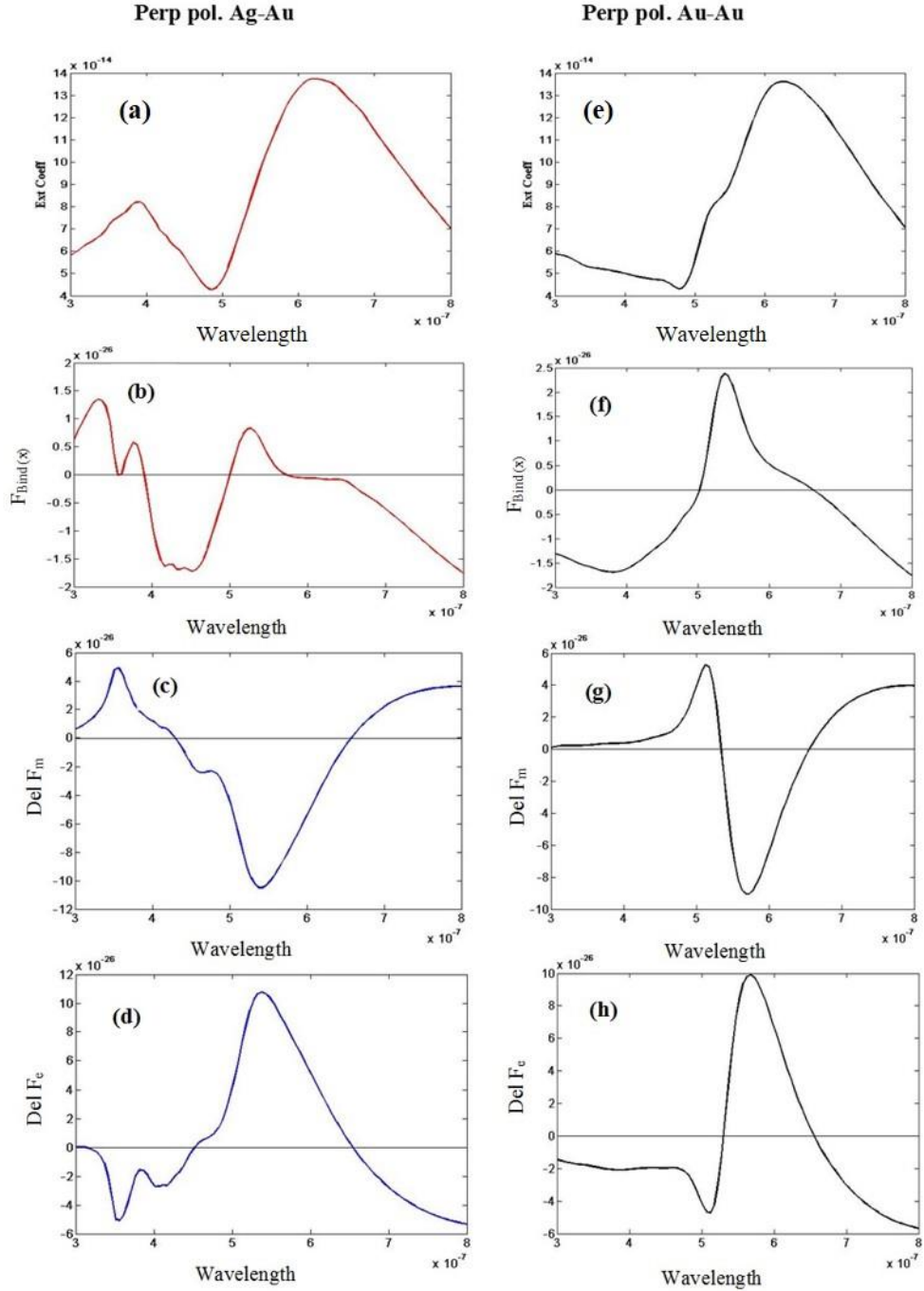

Fig.3: Considering perpendicular polarized light for the configuration of Fig.1 (a) and '$\varphi$' = 0 degree [on axis Ag-Au]: (a) Extinction co-efficient. (b) The lateral binding force $F_{Bind(x)} = (F_{B(x)} - F_{S(x)})$. (c) Difference of bulk Lorentz force. (d) Difference of surface force. Considering same polarization of light for the configuration of Fig.1 (b) and '$\varphi$' = 0 degree [on axis Au-Au]: (e) Extinction co-efficient (f) The lateral binding force $F_{Bind(x)} = (F_{B(x)} - F_{S(x)})$ (g) Difference of Lorentz bulk force. (h) Difference of Lorentz surface force.

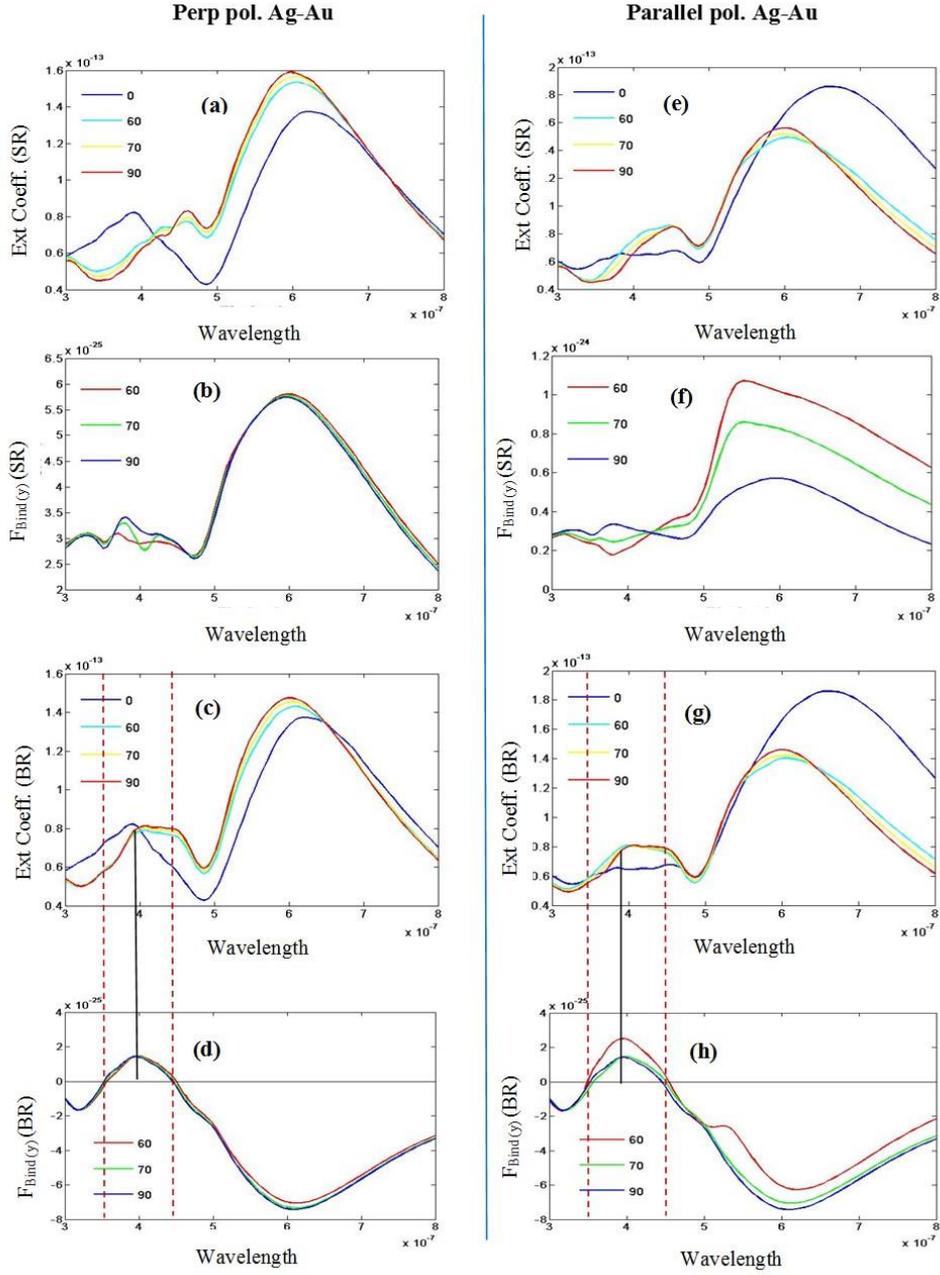

Fig.4: SR and BR represent 'small rotate' and 'big rotate' respectively and '$\varphi$' = 60, 70 and 90 degree [off axis Ag-Au]. Considering perpendicular polarized light- for the configuration of Fig.1 (a): (a) Extinction co-efficient (SR) (b) The longitudinal binding force $F_{Bind\,(y)}$ (SR); and for the configuration of Fig.1 (d): (c) Extinction co-efficient (BR) (d) the longitudinal binding force $F_{Bind\,(y)}$ (BR). Considering parallel polarized light: for the configuration of Fig.1 (a): (e) Extinction co-efficient (SR) (f) The longitudinal binding force $F_{Bind\,(y)}$ (SR); and for the configuration of Fig.1 (d): (g) Extinction co-efficient (BR) (h) The longitudinal binding force $F_{Bind\,(y)}$ (BR).

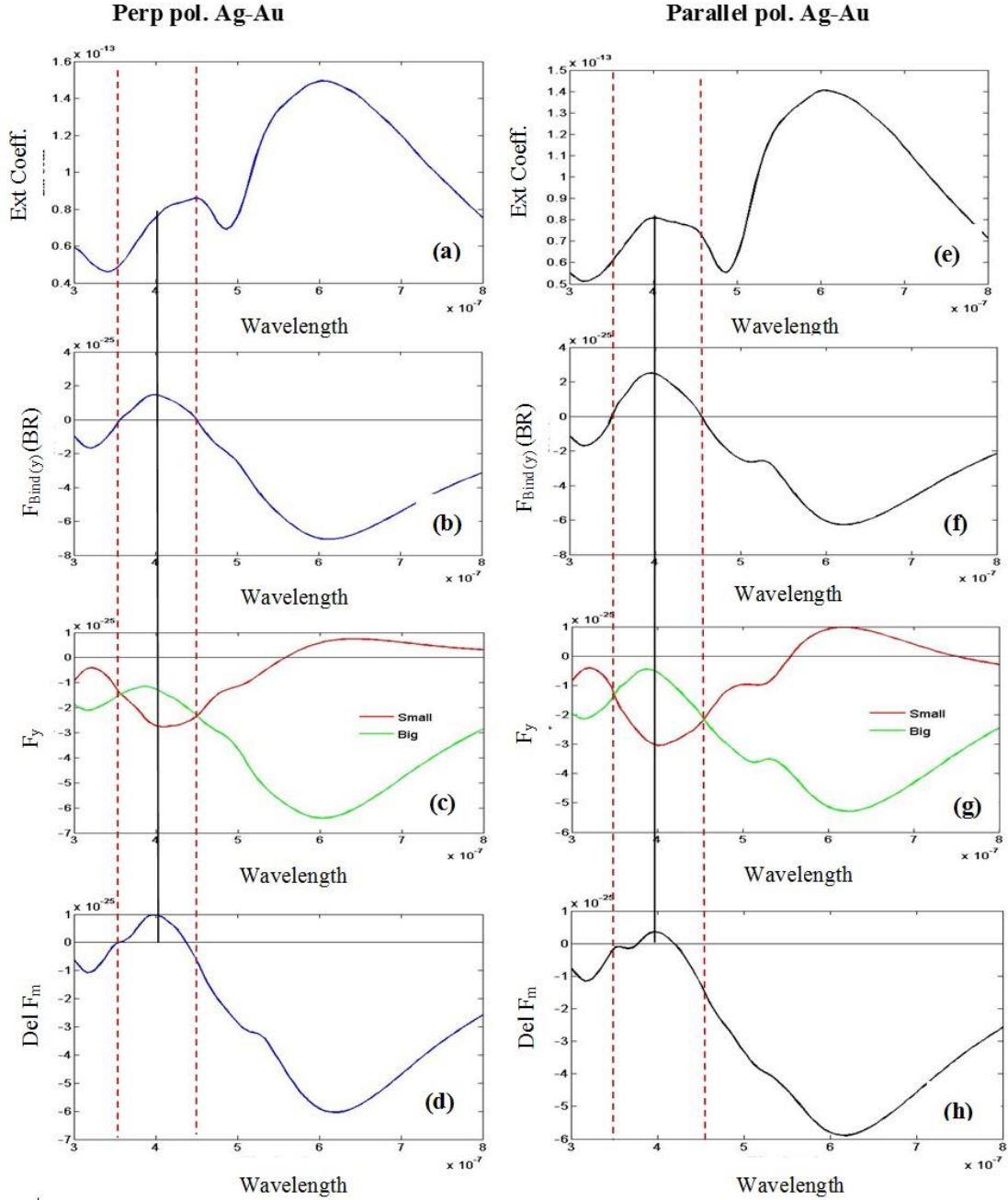

Fig.5: SR and BR represent 'small rotate' and 'big rotate' respectively. For off axis Ag-Au and BR [the configuration of Fig.1 (d) and '$\varphi$' = 60 degree]: Considering perpendicular polarized light (a) Extinction co-efficient (b) The longitudinal binding force $F_{Bind\ (y)}$ (BR) (c) Time averaged force on each particle. (d) Difference of bulk Lorentz force. Considering parallel polarized light for same configuration (e) Extinction co-efficient (f) The longitudinal binding force $F_{Bind\ (y)}$ (BR) (g) Time averaged force on each particle. (h) Difference of bulk Lorentz force.

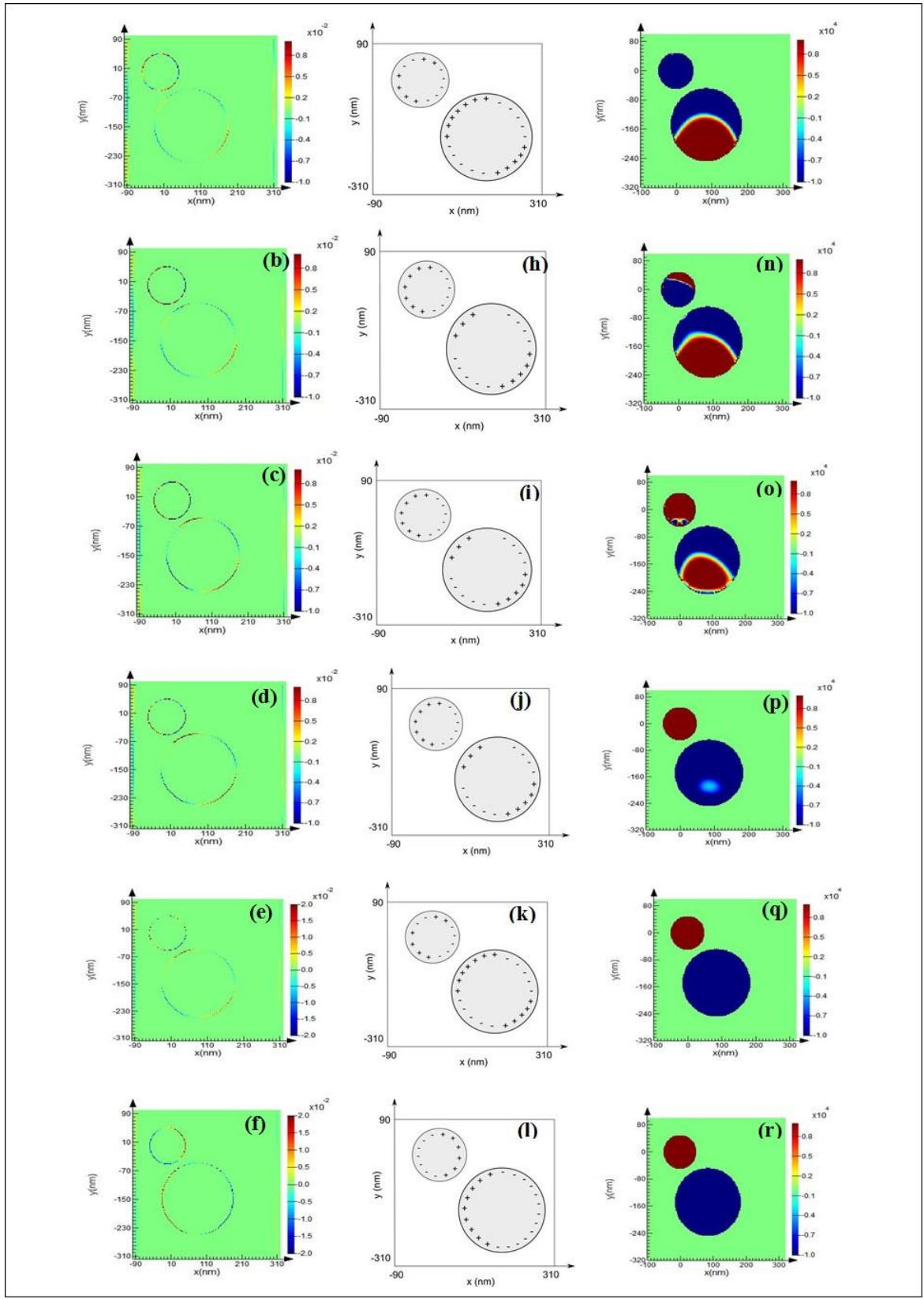

Fig. 6: For off axis Ag-Au and by rotating the big particle [the configuration of Fig.1 (d) and '$\varphi$' = 60 degree]: Considering parallel polarized light, from left first two columns represent surface charges

[(a)-(l)] and the third column represents steady state current [(m)-(r)]. We have chosen six wavelengths for six different rows (from top to bottom): 338, 354, 400, 457, 485 and 612 nm. Charge distributions: (a) QQ (b) DQ (c) DQ (d) DQ (e) QQ (f) DD; where Q and D mean quadrupole and dipole respectively.